\documentclass[runningheads]{llncs}
\usepackage[T1]{fontenc}
\usepackage{graphicx}
\usepackage[usenames,dvipsnames]{xcolor}
\usepackage{todonotes} \presetkeys{todonotes}{inline}{}
\usepackage{hyperref}
\usepackage{relsize}
\usepackage{xspace}
\usepackage[numbers]{natbib}
\usepackage{blindtext}
\usepackage{booktabs}
\usepackage[flushleft]{threeparttable}
\usepackage{tabularx}
\usepackage{rotating}
\usepackage{siunitx}
\usepackage[capitalise]{cleveref}
\usepackage{upgreek}

\newcommand{\eg}{e.g.\ }

\newcommand{\CPP}{\mbox{C\hspace{-.05em}\raisebox{.4ex}{\relsize{-3}{\textbf{++}}}}\xspace}
\newcommand{\ilqgamesjl}{{iLQGames.jl}\xspace}
\newcommand{\result}[3]{% args mean, sem, color, len
    $\SI{#1}{ms}$ & {\color{green!#2!red}\rule{#3cm}{8pt}}
}
\newcommand{\noresult}{ n/a& }

\begin{document}
\title{\ilqgamesjl: Rapidly Designing and Solving Differential Games in Julia}
%\titlerunning{Abbreviated paper title}
\author{
Lasse Peters \inst{1}\orcidID{0000-0001-9008-7127} \and
Zachary N. Sunberg\inst{2}\orcidID{0000-0001-9707-3035}
}
\authorrunning{L. Peters and Z. Sunberg}
% First names are abbreviated in the running head.
% If there are more than two authors, 'et al.' is used.
%
\institute{Hamburg University of Technology, Hamburg, Germany\\
\email{lasse.peters@tuhh.de} \and
University of Colorado Boulder, CO, USA\\
\email{zachary.sunberg@colorado.edu}}
\maketitle              % typeset the header of the contribution
\begin{abstract}

In many problems that involve multiple decision making agents, optimal choices for each agent depend on the choices of others.
Differential game theory provides a principled formalism for expressing these coupled interactions
and recent work offers efficient approximations to solve these problems to non-cooperative equilibria.
\ilqgamesjl is a framework for designing and solving differential games, built around the iterative linear-quadratic method presented in \cite{fridovich2019efficient}.
It is written in the Julia programming language to allow flexible prototyping and integration with other research software, while leveraging the high-performance nature of the language to
allow real-time execution.
The open-source software package can be found at \url{https://github.com/lassepe/iLQGames.jl}.

\end{abstract}

\section{Introduction and Related Work}

In order for a robot to be truly autonomous it needs to be capable of interacting with environments that are not isolated, but rather shared with other agents.
Naturally, these shared environments are manipulated by multiple decision making agents at a time and different agents may pursue different objectives.
For any non-trivial interaction scenario, the success of a given agent depends not only on its own actions but also on the decisions of others.
Therefore, agents must consider the effects of their actions on the behavior of others.

Dynamic game theory offers an expressive theoretical framework for formulating these types of interaction problems.
Differential games are a special class of dynamic games that consider problems in continuous domain.
In this framework, the evolution of the state is characterized by a differential equation which depends upon each players input, and the objectives of all players are expressed by their respective cost functions.
The solution of a differential game to an appropriate equilibrium concept, \eg a Nash equilibrium in the case of non-cooperative scenarios, provides a strategy assignment for \emph{all} players.
Game solutions may be used for centralized control of multiple agents or for (decentralized) control of an individual agent interacting with others \cite{fisac2019hierarchical,wang2019driving,peters2020inference}.
Recent work has demonstrated the effectiveness of game theoretic planning approaches in several multi-agent problems, such as human-robot interaction for intersection and highway driving \cite{fisac2019hierarchical,fridovich2019efficient,peters2020inference} or multi-robot racing \cite{spica2018real,wang2019driving,wang2019racing}.

Unfortunately, the complexity involved with describing and solving these problems impedes application to scenarios characterized by high-dimensional states, real-time constraints and fast planning rates, such as robotics.
Here, the term \emph{complexity} is explicitly used with threefold meaning:
First, the algorithmic time and space complexity of solution methods;
Second, the challenges involved with implementing these algorithms efficiently;
And third, the conceptual complexity of the interface used to describe and set up the problem.
While the first two aspects are crucial for quick solution of the problem, we emphasize that the latter aspect is particularly important to admit quick iteration of different designs, \eg to experiment with different cost structures that encode the behavior of each player.

In terms of computational complexity recent work offers efficient approximations to non-cooperative games \cite{fridovich2019efficient,spica2018real} and several works have demonstrated real-time performance of these algorithms in \CPP implementations \cite{fridovich2019efficient,spica2018real,wang2019driving,wang2019racing}.
However, to the best of our knowledge, only \cite{fridovich2019efficient} provide a publicly available implementation of their solver\footnote{\url{https://github.com/HJReachability/ilqgames}} and little work has focused on providing flexible interfaces and tools for the design phase of differential games.

This work presents \ilqgamesjl, a framework for designing and solving differential games using the iterative linear-quadratic (iLQ) method proposed in \cite{fridovich2019efficient}.
\ilqgamesjl is written in the Julia programming language \cite{bezanson2012julia} and makes use of the language's genericity to provide a flexible interface that admits quick iteration of different problem designs and keep up with execution times of a comparable \CPP implementation.
This paper describes the key aspects of the framework that enable its flexibility and performance, and make it an effective tool for differential game research.

\section{Architecture for Rapid Design and Solution}

\subsection{Rapid Design}

When modelling a practical scenario of multi-agent interaction as a differential game, it may not be immediately clear what are suitable dynamics and costs to describe the problem.
Therefore, \ilqgamesjl provides a thin interface for describing differential games that allows users to set up a model in few lines of code.
\footnote{See \url{https://github.com/lassepe/iLQGames.jl} for a code example.}

\begin{figure}
  \centering
  \includegraphics[width=0.9\linewidth]{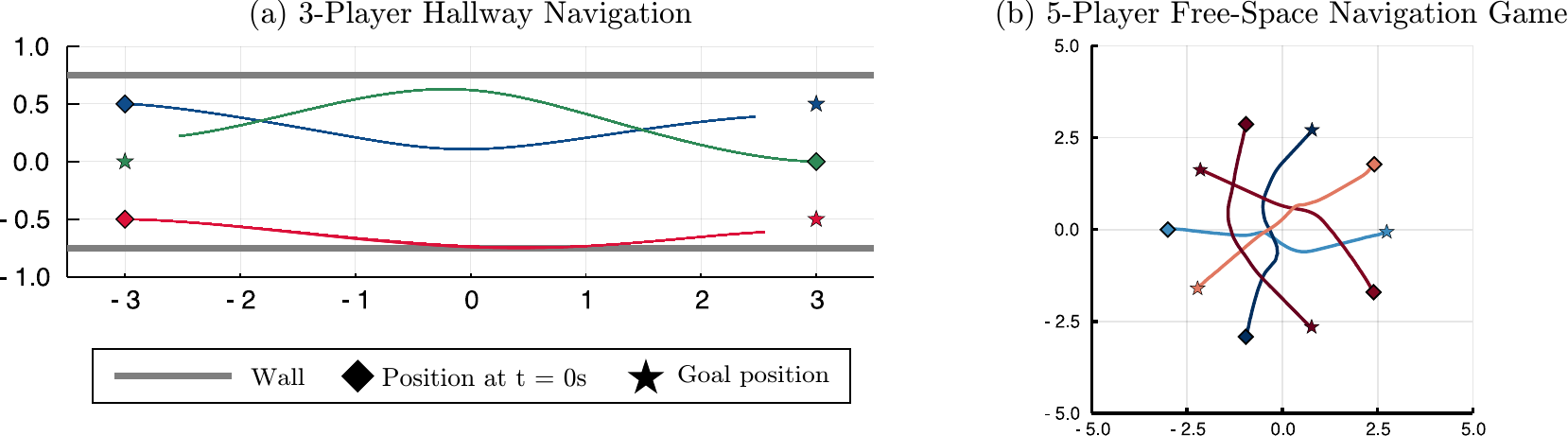}
  \caption{
  Two differential games solved with \ilqgamesjl: a 3-player collision avoidance problem in a hallway and a 5-player collision avoidance problem in free-space.
  }%
  \label{fig:example}
\end{figure}

Two examples of differential games that are designed, solved and visualized using \ilqgamesjl are given in \cref{fig:example}.
In these problems, each player is in control of a subsystem with 4D unicycle dynamics and the objective of each agent is to reach their respective goal while avoiding collisions with the wall or other agents.
After providing the differential equation characterizing the game dynamics (here, a product system composed of multiple unicycle subsystems) and specifying the cost functions for each player (here, penalizing proximity, final distance to the goal location and control effort), the user can directly invoke the iLQ solver.
Most notably, even though the iLQ solver is based on successive linear-quadratic (LQ) approximations of the game, the user does \emph{not} need to hand-specify partial derivatives for linearization of the dynamics or quadratization of the costs. Instead, \ilqgamesjl can compute LQ approximations efficiently via automatic differentiation of the game dynamics and player costs using \cite{RevelsLubinPapamarkou2016}.

Furthermore, being written in pure Julia ---~a language with strong focus on scientific programming~--- \ilqgamesjl can directly be used with various other packages from the ecosystem.
%For example, \ilqgamesjl natively supports visualization of the state and input trajectories of a game solution (\cf \cref{fig:example}) or projections of the cost landscape for a given player to support the design process.
An example that demonstrates this advantage is presented in \cite{peters2020inference}, where \ilqgamesjl is combined with ParticleFilters.jl to reason about behavioural uncertainty of other players in differential games.

\subsection{Rapid Solution}

Despite being a high-level language that offers wide-ranging abstraction, the Julia compiler generates highly optimized code.
% TODO: prunable
%\footnote{This has been prominently demonstrated in \cite{regier2019cataloging}.}
Most notably for our use case, this allows \ilqgamesjl to solve each LQ iterate very efficiently via a fully stack-allocated dynamic program implemented in a readable high-level style in less than 70 lines of code.
In fact, for moderately sized games, this optimization allows our implementation to outperform the \CPP implementation presented in \cite{fridovich2019efficient}.
% TODO: prunable
%While it must be noted that \cite{fridovich2019efficient} does not make this optimization and in theory a \CPP implementation can achieve similar performance, we argue that making comparable optimizations in \CPP may not be possible without sacrificing readability.

Furthermore, the generic function dispatch mechanism used in Julia allows users to overload default implementations at various levels of the solver to make problem specific optimizations.
For example, users can specify a custom method to perform LQ approximations once they have chosen a design for the problem.

\begin{sloppypar}
Beyond that, \ilqgamesjl supports exploitation of special structure of the dynamics to speed up computation.
This aspect is realized via the \texttt{LinearizationStyle} trait concept.
By default, a dynamical system is attributed the \texttt{JacobianLinearization} trait and automatic differentiation or a user-defined linearization is used to obtain LQ approximations of the game.
However, if the dynamics are feedback linearizable, the user can optionally specify the \texttt{FeedbackLinearization} trait for a model to invoke a specialized version of the solver presented in \cite{fridovich2019flat}.
% TODO: prunable
% Finally, linear systems are assigned the \texttt{TrivialLinearization} trait and linearization is explicitly skipped.
This trait concept can be easily extended to other special types of systems and thus allows users to seamlessly customize the solver with small local changes without the need to overload other parts of the routine.
\end{sloppypar}

\section{Performance}

The performance of \ilqgamesjl is evaluated by benchmarking it on three problems against the \CPP implementation presented in \cite{fridovich2019efficient}.
For additional reference, a Python implementation of an LQ solver is benchmarked as well.
The benchmark problems are a minimal LQ game, a nonlinear nonquadratic collision avoidance problem similar to the examples depicted in \cref{fig:example}, and a feedback linearized version of the latter.

\Cref{tab:benchmark} summarizes the benchmark results.
The Python implementation for the LQ case is multiple orders of magnitude slower than the \CPP version and \ilqgamesjl and thus would not scale well to nonlinear nonquadratic problems.
When utilizing manually specified partial derivatives to compute LQ approximations, as also done in the \CPP version, \ilqgamesjl outperforms the baseline.
When using automatic differentiation \ilqgamesjl still achieves moderate runtime and is sufficiently fast to evaluate different problem designs.

\begin{table}
    \caption{Benchmark results. The tuple behind each problem indicates the number of players (P) and the dimensionality of the state (D).
    LQ-Python denotes a Python implementation of the dynamic program used at the inner loop of the iLQ game algorithm.
    iLQGames-\CPP refers to the implementation used in \cite{fridovich2019efficient}.
    \mbox{\ilqgamesjl-MD} and \mbox{\ilqgamesjl-AD} refer to our implementation, using manual differentiation and automatic differentiation, respectively.
    Each game is solved over a horizon of 100 time steps on a standard laptop.
    } \label{tab:benchmark}

    \begin{tabularx}{\linewidth}{lXrlXrlXrl}
        \toprule
        & & \multicolumn{2}{c}{LQ (2P, 2D)} & & \multicolumn{2}{c}{Nonlinear (3P, 12D)} & & \multicolumn{2}{c}{FBLinearized (3P, 12D)}\\
        \midrule
        LQ-Python       & & \result{20.800}{0}{1.0}      & & \noresult                & & \noresult \\
        iLQGames-\CPP   & & \result{0.3490}{98}{0.17}    & & \result{16.27}{84}{0.16} & & \result{13.25}{87}{0.13}  \\
        \ilqgamesjl-MD  & & \result{0.0044}{100}{0.0002} & & \result{7.19}{92}{0.072} & & \result{3.98}{96}{0.04}   \\
        \ilqgamesjl-AD  & & \noresult                    & & \result{63.57}{37}{0.63} & & \result{52.50}{47}{0.53} \\
        \bottomrule
    \end{tabularx}

\end{table}

\section{Conclusion}

\ilqgamesjl is a framework for designing and solving differential games, built around the iterative linear-quadratic method presented in \cite{fridovich2019efficient}.
This manuscript provides an overview of the framework and discusses key design aspects that enable its flexibility and performance.
\ilqgamesjl provides a first step towards making differential games an easily accessible tool for multi-agent interaction research.
Future directions include abstraction of a high-level problem interface that can be shared between multiple solvers to simplify the process of benchmarking algorithms against one another.

%  \section*{Acknowledgements}
%  
%  The authors would like to thank David Fridovich-Keil for helpful discussions on iLQ games and their efficient realization.

% ---- Bibliography ----
%
% BibTeX users should specify bibliography style 'splncs04'.
% References will then be sorted and formatted in the correct style.
%
\bibliography{main.bib}

\end{document}